\documentclass[a4paper]{aa}
\def\Om{{\it \Omega}}

\usepackage{psfig}

\renewcommand{\vec}[1]{\mbox{\boldmath $#1$}}
\def\qq{\qquad\qquad}                      
\def\qqq{\qquad\qquad\qquad}               
\def\q{\qquad}
\def\beg{\begin{eqnarray}}
\def\ende{\end{eqnarray}}
\def\Om{{\it \Omega}}
\begin{document}

\title{Stability of density-stratified viscous Taylor-Couette flows}
\titlerunning{Stability of density-stratified viscous Taylor-Couette flows}
\author{D. Shalybkov\inst{1,2} \and G. R\"udiger\inst{1,3}}

\institute{
Astrophysikalisches Institut Potsdam,
         An der Sternwarte 16, D-14482 Potsdam, Germany
\and A.F. Ioffe Institute for Physics
and Technology, 194021, St. Petersburg, Russia \and Isaac Newton Institute for Mathematical Sciences, 20 Clarkson Road, Cambridge, CB3 0EH, U.K.}

\date{\today}

\abstract{
The stability of density-stratified viscous Taylor-Couette flows is considered
using the Boussinesq approximation but without any use of the short-wave
approximation.
The   flows which are unstable after the Rayleigh
criterion ($\hat \mu<\hat \eta^2$, with  $\hat \mu=\Om_{\rm out}/\Om_{\rm in }$ and 
$\hat \eta=R_{\rm in}/R_{\rm out}$) now develop overstable axisymmetric
Taylor vortices.  For the considered wide-gap container we find  the nonaxisymmetric modes  as  the most
unstable ones.  
The nonaxisymmetric
modes are unstable also {\em beyond the Rayleigh line}. For such modes  the instability condition seems simply  to be 
$\hat\mu<1$  as stressed  by  Yavneh, McWilliams \& Molemaker 
(2001).  However, we never found unstable modes  for too flat rotation laws
fulfilling the condition  $\hat \mu>\hat \eta$. The Reynolds numbers  rapidly grow  to very high values
if this limit is approached (see Figs. \ref{remu} and \ref{remu2}).
 Also striking is that the marginal stability lines for the higher $m$  do less and less enter the region beyond the  Rayleigh line  so that   we might have to  consider the stratorotational instability   as a  'low-$m$ instability'. 
\\
The applicability of these results to the stability problem of accretion
disks  with their strong stratification and fast rotation  is shortly discussed. 
}

\maketitle
\section{Introduction}
The  flow pattern between concentric rotating cylinders 
with a stable axial density stratification was firstly studied by Thorpe
(1966) who concluded that stable stratification stabilizes the flow.
The further experimental and theoretical studies by Boubnov, Gledzer \&
Hopfinger (1995) 
 and Caton, Janiaud \& Hopfinger (2000) confirmed the stabilizing role of the
density-stratification and showed that i) the critical Reynolds number
depend on the buoyancy frequency (or Brunt-V\"ais\"al\"a frequency) of the fluid
and ii) the stratification reduces the vertical extension of the Taylor vortices.
The computational results of Hua, Le Gentil \& Orlandi (1997) have indeed 
reproduced the experiment results. 

The common feature of these studies is that the outer
cylinder is at rest and flow is unstable after the Rayleigh condition for inviscid
flow (which was extended to stratified fluids by Ooyama 1966), i.e.
\beg
\frac{{\rm d}}{{\rm d}R}( R^4\Om^2) <0
\label{ray}
\ende
where $\Om$ is the angular velocity of the flow.

Recently, Molemaker, McWilliams \& Yavneh (2001)
and Yavneh, McWilliams \& Molemaker (2001)  found 
\beg
\frac{{\rm d} \Om^2}{{\rm d}R}<0
\label{mri}
\ende
as the sufficient condition for (nonaxisymmetric) instability.
The condition (\ref{mri}) is identical with the condition for magnetorotational
instability of Taylor-Couette flow (Velikhov 1959).
These  results have been derived by a linear stability analysis for inviscid flow.
The numerical results of Yavneh, McWilliams \& Molemaker (2001)
demonstrate the existence of the hydrodynamic  instability also for finite viscosity.
\begin{figure}[ht]
\psfig{figure=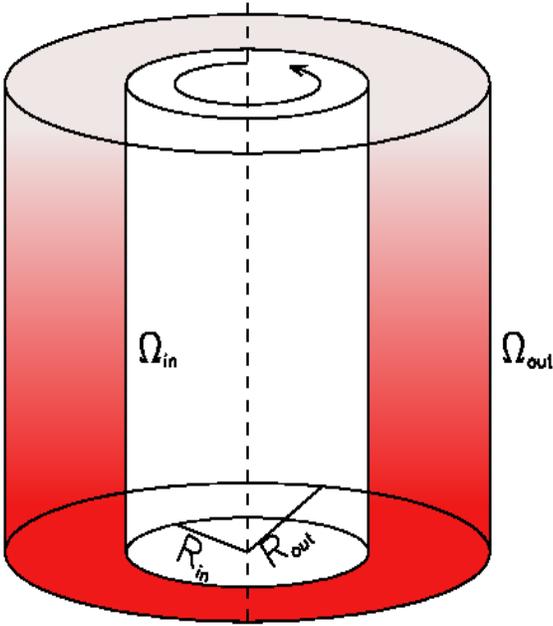,width=8cm,height=9.0cm} 
\caption{\label{cont} The geometry of density-stratfied Taylor-Couette experiments.}
\end{figure}

An instability of  density-stratified Taylor-Couette flow  beyond the Rayleigh line  $\hat\mu=\hat\eta^2$
for  nonaxisymmetric disturbances have  also been  found experimentally (Withjack \& Chen 1974).
With their wide gap container ($\hat\eta=0.2$)  they  found for $\hat \mu>0$  the 
stability curve {\it crossing} the classical Rayleigh line. The observed instability was reported as {\it nonaxisymmetric}.
The resulting experimental stability line, however,  is very steep
for positive $\hat\mu$ (see their  Fig. 8)
and does never cross the line $\hat\mu=\hat\eta$.   

 For real viscous flows there are very  illustrative results by Yavneh, McWilliams \& Molemaker (2001).  In the present paper a more comprehensive study
of such  flows is given. The governing equations and the restrictions of the  used Boussinesq approximation
are  discussed in Sect.~2 while the numerical results are presented in
Sect.~\ref{res}. Summary and final discussion are given in Sect.~\ref{disc}.

The existence of an instability in  Taylor-Couette flows with a stable 
radial rotation law and with a stable $z$-stratification of the density 
is a surprise in the light of the Solberg-H\o{}iland criterion 
(see R\"udiger, Arlt \& Shalybkov 2002). The necessary condition for stability
reads 
\beg 
{\partial g_R \over \partial z} = {\partial g_z \over \partial R}, 
\label{eg} 
\ende 
with $\vec{g}$ is the external force acceleration. 
Any {\em conservative} force is 
a particular solution of (\ref{eg}). Without external forces this relation
is thus always fulfilled.  If -- as it is in accretion disks -- the gravity  
balances the pressure and the centrifugal force, then Eq. (\ref{eg}) is 
automatically fulfilled.  
Note that after the Poincare theorem for rotating media with potential force
and $\Om=\Om(R)$ both the density and the pressure can be written as 
functions of the generalized potential so that (\ref{eg}) is always fulfilled. 
 Generally, the magnetic field is {\em not} conservative and can never 
fulfill the condition (\ref{eg}). This is the basic explanation for the 
existence of the magnetorotational instability  (MRI) driven by (weak) 
magnetic fields.  

Equation  (\ref{eg}) has been derived by means of the short-wave
approximation $m<R/\delta R$. For waves which are  large in radial
directions this condition might easily be fulfilled only for $m=0$. 
It is, therefore,  important to break the short-wave approximation
in order to probe also the  nonaxisymmetric modes.
For Kepler flows  (with uniform gravity acceleration)  this
has recently been done by Dubrulle et al. (2004). With  Boussinesq approximation and  direct numerical simulation for viscous flow they 
find  all stratified flows with negative ${\rm d}\Om/{\rm d}R$  unstable against nonaxisymmetric disturbances.
According to their results a critical
Froude number (as defined by Eq. \ref{Fr}) exists below which the flow is stable.  Whether the Boussinesq approximation can be used for too small  Froude numbers seems still to be an open question.  

\section{Equations and basic state}
In cylindric coordinates ($R$, $\phi$, $z$) the equations of
incompressible stratified fluid with  uniform dynamic viscosity,
$\mu$, are
\beg
\lefteqn{\frac{\partial u_R}{\partial t}+(\vec{u}\nabla)u_R-\frac{u_\phi^2}{R}
 = } \nonumber \\
&& \quad\quad\quad\quad -\frac{1}{\rho}\frac{\partial P}{\partial R} 
+ \nu \left[ \Delta u_R - \frac{2}{R^2}\frac{\partial u_\phi}{\partial
\phi} - \frac{u_R}{R^2} \right], \nonumber \\
\lefteqn{ \frac{\partial u_\phi}{\partial t}+(\vec{u}\nabla )u_\phi+
\frac{u_\phi u_R}{R} =}  \nonumber \\
&& \quad\quad\quad\quad -\frac{1}{\rho R}\frac{\partial P}{\partial \phi} 
+ \nu \left[ \Delta u_\phi + \frac{2}{R^2}\frac{\partial u_R}{\partial
\phi} - \frac{u_\phi}{R^2} \right], \nonumber \\
\lefteqn{\frac{\partial u_z}{\partial t}+(\vec{u}\nabla )u_z=
-\frac{1}{\rho }\frac{\partial P}{\partial z} - g 
+ \nu \Delta u_z,} \nonumber \\
\lefteqn{\frac{\partial u_R}{\partial R}+\frac{u_R}{R}+
\frac{1}{R} \frac{\partial u_\phi}{\partial \phi}+\frac{\partial u_z}
{\partial z}=0,} 
\label{sys}
\ende
where
\beg
(\vec{u} \nabla)u_R=u_R\frac{\partial u_R}{\partial R}+\frac{u_\phi}{R}
\frac{\partial u_R}{\partial \phi}+u_z\frac{\partial u_R}{\partial z}
\label{unabla}
\ende
and
\beg
\Delta u_R=\frac{\partial^2 u_R}{\partial R^2}+\frac{1}{R}
\frac{\partial u_R}{\partial R}+\frac{1}{R^2}\frac{\partial^2 u_R}{\partial
\phi^2}+\frac{\partial^2 u_R}{\partial z^2}.
\ende
$\rho$ is
the density, $P$ is the pressure, $g$ is the gravity,
$\nu=\mu/\rho$ is the kinematic viscosity.\footnote{
the density diffusion term in the mass conservation equation is neglected  (see e.g. Caton, Janiaud \&
Hopfinger 2000)} The equation which describes the evolution  of the density fluctuation moving in the general density field is
\beg
\frac{\partial \rho}{\partial t} + (\vec{u} \nabla)\rho =0.
\label{density}
\ende
We have to formulate  the basic state with prescribed  
velocity profile $\vec{u}=(0,R\Om(R),0)$ and given density
vertical stratification $\rho=\rho(z)$.
The system (\ref{sys}) takes the form
\beg
\lefteqn{ \frac{u_\phi^2}{R}=\frac{1}{\rho}\frac{\partial P}{\partial R} ,
\quad\quad\quad\quad\quad
\frac{1}{\rho }\frac{\partial P}{\partial z} = - g,}
\nonumber \\
\lefteqn{ \frac{\partial^2 u_\phi}{\partial R^2}+\frac{1}{R}
\frac{\partial u_\phi}{\partial R} - \frac{u_\phi}{R^2} =0.}
\label{sysb}
\ende
The last equation defines the angular velocity
\beg
\Om=a+\frac{b}{R^2},
\label{Om}
\ende
where $a$ and $b$ are two constants related to the boundary values of
the angular velocity, $\Om_{\rm in}$, $\Om_{\rm out}$, of the
inner cylinder with radius, $R_{\rm in}$, and the outer
cylinder with radius, $R_{\rm out}$. It follows           
\beg
a=\Om_{\rm in}\frac{\hat \mu - \hat \eta^2}{1-\hat\eta^2}, \qq
b=\Om_{\rm in}R_{\rm in}^2\frac{1-\hat\mu}{1-\hat\eta^2},
\label{ab}
\ende
with
\beg
\hat\mu =\frac{\Om_{\rm out}}{\Om_{\rm in}}, \qq
\hat\eta =\frac{R_{\rm in}}{R_{\rm out}}.
\ende
Differentiating the first equation of the system (\ref{sysb}) by $z$
and the second equation by $R$, subtracting each other and using the
supposed profiles of density and angular velocity
one gets
\beg
R\Om^2 \frac{{\rm d} \rho}{{\rm d} z} =0.
\label{cond}
\ende
After this relation the density can depend only on the vertical
coordinate $z$ in the absence of rotation ($\Om=0$) and the angular
velocity can only depend on radius in the absence of the vertical
density stratification (${\rm d}\rho/{\rm d}z=0$). The supposed profiles
of the angular velocity, $\Om=\Om(R)$, and the density $\rho=\rho(z)$
are, therefore, not self-consistent. 
Thus, we must admit more 
general profile for the density  $\rho=\rho(R,z)$ even though the initial
stratification for the resting fluid is only vertical. In this case,
the condition (\ref{cond}) takes the form
\beg
R\Om^2 \frac{\partial \rho}{\partial z}
+g\frac{\partial \rho}{\partial R} =0.
\label{cond1}
\ende
The fluid transforms under the centrifugal force from
the pure vertical stratification at the initial state to 
mixed (vertical and radial) stratification under the influence of the rotation
strongly complicating  the problem. 

For real experiments the initial (without rotation)
vertical stratification is small $|d{\textrm{log}}\rho/d
{\textrm{log}}z \ll 1 |$  as is the ratio of centrifugal acceleration to
 the vertical gravitation acceleration 
\beg
\left|\frac{R^2\Om}{g}\right| \ll 1,
\label{conda}
\ende
so that after (\ref{cond1}) the radial stratification is also small.
Let us therefore consider the case of a weak stratification 
\beg
\rho=\rho_0+\rho_1(R,z), \q \rho_1 \ll \rho_0,
\label{smrho}
\ende
where $\rho_0$ is the uniform background density and 
(\ref{cond1}) is fulfilled in zero-order.
The perturbed state of the flow is described by
\beg
\lefteqn{ u_R, \qqq u_\phi+R\Om(R), \qqq u_z,} 
\nonumber \\
\lefteqn{ P_0(R)+P_1(R,z)+P, \qq \rho_0+\rho_1(R,z)+\rho,}
\ende
where $|P_1/P_0| \ll 1$ and $u_R$, $u_\phi$, $u_z$, $P$ and
$\rho$ are the perturbations.
Linearizing the system (\ref{sys}) and selecting only the terms of
the largest order we have
\beg
\lefteqn{ \frac{\partial u_R}{\partial t} + \Om \frac{\partial u_R}{\partial
\phi}- 2\Om u_\phi=-\frac{1}{\rho_0}\frac{\partial P}{\partial R}
+\frac{\rho}{\rho_0^2}\frac{\partial P_0}{\partial R} +}\nonumber \\ 
&& \quad\quad\quad\quad + \nu_0 \left[ \nabla u_R-\frac{2}{R^2}\frac{\partial u_\phi}{\partial \phi}-
\frac{u_R}{R^2} \right], \nonumber \\
\lefteqn{ \frac{\partial u_\phi}{\partial t} + \Om \frac{\partial u_\phi}{\partial
\phi}+\frac{1}{R}\frac{\partial }{\partial R}(R^2 \Om)u_R=
-\frac{1}{\rho_0 R}\frac{\partial P}{\partial \phi} +}\nonumber \\ 
&& \quad\quad\quad\quad + \nu_0 \left[ \nabla u_\phi+\frac{2}{R^2}\frac{\partial u_R}{\partial \phi}-
\frac{u_\phi}{R^2} \right], \nonumber \\
\lefteqn{ \frac{\partial u_z}{\partial t} + \Om \frac{\partial u_z}{\partial
\phi}=-\frac{1}{\rho_0}\frac{\partial P}{\partial z}
+\frac{\rho}{\rho_0^2}\frac{\partial P_0}{\partial z}+ \nu_0 \nabla u_z,}
\nonumber \\
\lefteqn{ \frac{\partial \rho}{\partial t}+\frac{\partial \rho_1}{\partial R}u_R
+\Om \frac{\partial \rho}{\partial \phi}+\frac{\partial \rho_1}
{\partial z}u_z=0,} \nonumber \\
\lefteqn{ \frac{\partial u_R}{\partial R}+\frac{u_R}{R}+\frac{1}{R}
\frac{\partial u_\phi}{\partial \phi}+\frac{\partial u_z}{\partial z}
=0}
\label{sysl}
\ende
with uniform $\nu_0=\mu /\rho_0$. 
The first-order terms are left in the mass conservation equation due
to vanishing of the zero-order terms.

Due to (\ref{conda}) we can neglect $\partial P_0/\partial R$ in the first
equation and $\partial\rho_1/\partial R$ in the fourth arising from radial
stratification (they will be $|R^2\Om /g|$ times smaller than
terms arising from the vertical stratifications)
and the system takes exactly the Boussinesq form
\beg
\lefteqn{ \frac{\partial u_R}{\partial t} + \Om \frac{\partial u_R}{\partial
\phi}- 2\Om u_\phi=
-\frac{\partial }{\partial R}\left(\frac{P}{\rho_0}\right)+}
\nonumber \\ 
&& \quad\quad\quad\quad + \nu_0 \left[ \nabla u_R-\frac{2}{R^2}\frac{\partial u_\phi}{\partial \phi}-
\frac{u_R}{R^2} \right], \nonumber \\
\lefteqn{ \frac{\partial u_\phi}{\partial t} + \Om \frac{\partial u_\phi}{\partial
\phi}+\frac{1}{R}\frac{\partial R^2 \Om}{\partial R}u_R=
-\frac{1}{R}\frac{\partial }{\partial \phi}\left(\frac{P}{\rho_0}\right)+}
\nonumber \\ 
&& \quad\quad\quad\quad + \nu_0 \left[ \nabla u_\phi+\frac{2}{R^2}\frac{\partial u_R}{\partial \phi}-
\frac{u_\phi}{R^2} \right], \nonumber \\
\lefteqn{ \frac{\partial u_z}{\partial t} + \Om \frac{\partial u_z}{\partial
\phi}=-\frac{\partial }{\partial z}\left(\frac{P}{\rho_0}\right)
-g\frac{\rho}{\rho_0} + \nu_0 \nabla u_z,}
\nonumber \\
\lefteqn{ \frac{\partial }{\partial t}\left(\frac{\rho}{\rho_0}\right)
+\Om \frac{\partial }{\partial \phi}\left(\frac{\rho}{\rho_0}\right)
-\frac{N^2}{g}u_z=0,} \nonumber \\
\lefteqn{ \frac{\partial u_R}{\partial R}+\frac{u_R}{R}+\frac{1}{R}
\frac{\partial u_\phi}{\partial \phi}+\frac{\partial u_z}{\partial z}
=0,}
\label{sysbo}
\ende
where $N$ is the vertical buoyancy frequency with
\beg
N^2=-\frac{g}{\rho_0}\frac{\partial \rho_1}{\partial z}.
\ende
Suppose that the linear vertical density stratification
$\partial \rho_1/\partial z={\textrm{const}}$ and thus
$N^2$ is a constant, too. 
Then the coefficients of the system (\ref{sysbo}) only depend on the
radial coordinate and we can use a normal mode expansion of the solution
${F}={F}(R){\textrm{exp}}({\textrm{i}}(m\phi+kz+\omega t))$ where $F$
represents any of the disturbed quantities.

Let $D=R_{\rm out} - R_{\rm in}$ be the gap between
the cylinders. We use $R_0=(R_{\rm in}D)^{1/2}$
as the unit of length,
the velocity $\Om_{\rm in} R_0$ as the unit of
the perturbed velocity, $\Om_{\rm in}$ as the unit of
$\omega$, $N$ and $\Om$.
Using the same symbols for normalized quantities and redefining $\rho$
as the dimensionless density $\rho g/\rho_0R_0\Om_{\rm in}^2$
we finally find
\beg
\lefteqn{ \frac{\partial^2 u_R}{\partial R^2}
+\frac{1}{R}\frac{\partial u_R}{\partial R}
-\frac{u_R}{R^2} -\left(k^2+\frac{m^2}{R^2}\right) u_R
-2{\textrm{i}}\frac{m}{R}u_\phi-}
\nonumber \\
&& \quad\quad\quad\quad -{\textrm{i Re}}(\omega+m\Om) u_R
+2{\textrm{Re}}\Om u_\phi
-{\textrm{Re}}\frac{\partial P}{\partial R}=0,
\nonumber \\
\lefteqn{ \frac{\partial^2 u_\phi}{\partial R^2}
+\frac{1}{R}\frac{\partial u_\phi}{\partial R}
-\frac{u_\phi}{R^2} -\left(k^2+\frac{m^2}{R^2}\right) u_\phi
+2{\textrm{i}}\frac{m}{R}u_R-}
\nonumber \\
&& \quad -{\textrm{i Re}}(\omega+m\Om) u_\phi
-{\textrm{i Re}}\frac{m}{R}P
-\frac{{\textrm{Re}}}{R}\frac{\partial}{\partial R}(R^2 \Om)=0,
\nonumber \\
\lefteqn{ \frac{\partial^2 u_z}{\partial R^2}
+\frac{1}{R}\frac{\partial u_z}{\partial R}
-\left(k^2+\frac{m^2}{R^2}\right) u_z-}
\nonumber \\
&& \quad\quad\quad\quad -{\textrm{i Re}}(\omega+m\Om) u_z
-{\textrm{i Re}}\,kP-{\textrm{Re}}\,\rho=0,
\nonumber \\
\lefteqn{ {\textrm{i}}(\omega+m\Om)\rho
-N^2u_z=0}
\label{sysf}
\ende
and
\beg
 \frac{\partial u_R}{\partial R}+\frac{u_R}{R}
+{\textrm{i}}\frac{m}{R}u_\phi+{\textrm{i}}ku_z=0
\label{sysf1}
\ende
with the Reynolds  number 
\beg
{\textrm{Re}}=\frac{\Om_{\rm in}R_{\rm in} D}{\nu}.
\ende

The standard no-slip boundary conditions used at the inner and outer cylinder, i.e.
\beg
u_R=u_\phi=u_z=0,
\label{bound}
\ende
complete the classical eigenvalue problem.
The same numerical method as in our previous papers about the Taylor-Couette
problem (see e.g. R\"udiger \& Shalybkov 2002) is used. Here we use 
a small negative imaginary part of $\omega$ 
to avoid problems with the corotation point $\omega=m\Om$  for $m>0$.
Thus, the calculated critical Reynolds numbers are not for the
marginally stable state but for slightly unstable state.
To be sure that the calculated unstable state can be realized in experiments
we checked the existence of the transition from stable to unstable
state for several arbitrary points.
\begin{figure}[ht]
\psfig{figure=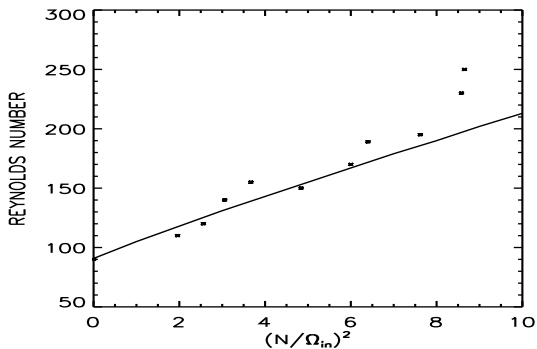,width=8cm,height=5.0cm} 
\caption{\label{comp} The marginal stability line for axisymmetric
disturbances ($m=0$) for  $\hat\eta=0.78$, $\hat\mu=0$. The dots
represent the experimental data of Boubnov, Gledzer \& Hopfinger (1995).}
\end{figure}

The code has been tested by computing  for $m=1$ the critical Reynolds number for the run 2 of Withjack \& Chen (1974, their Tab. 1) with the experimental  value 196.2 (with our normalizations)  and our  computed result 200.6 which we accepted  to be in sufficiently good  accordance.
\section{Results}
\label{res}
The imaginary parts of $\omega$, $\Im(\omega)$, decrease with increasing Reynolds number.
The Reynolds numbers above which the imaginary part of $\omega$
is smaller than some fixed value depend on the vertical wave number. They
have a minimum at a certain wave number for fixed other parameters.
This minimum value is called the critical Reynolds number.

In Fig.~\ref{comp} we compare the calculated marginal stability line
(i.e. $\Im(\omega)=0$) for axisymmetric disturbances
with experimental values by Boubnov, Gledzer \& Hopfinger (1995). The agreement
is rather good except
the small values of the Froude number
\begin{equation}
{\rm Fr}=\frac{\Om_{\rm in}}{N}.
\label{Fr}
\end{equation}
This disagreement 
may indicate the violation of the Boussinesq approximation. For Kepler disks we
find ${\rm Fr}\simeq 0.5$ (Dubrulle et al. 2004).  The unstratified fluids possess infinite Froude
number.
\begin{figure}[ht]
\psfig{figure=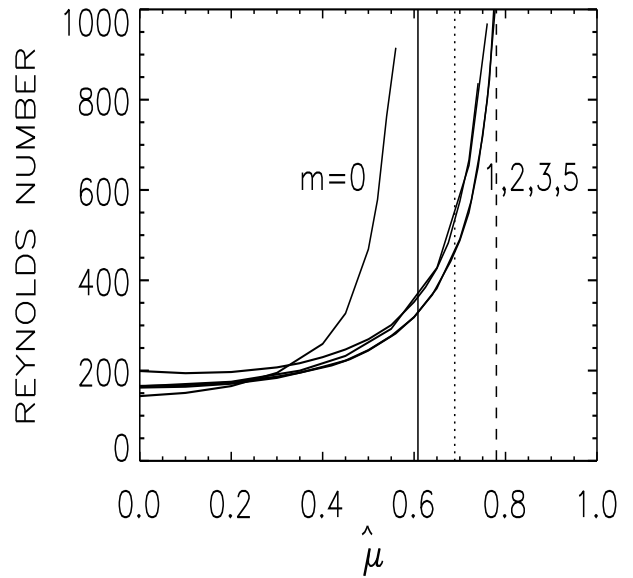,width=9.0cm,height=9.0cm} 
\caption{The marginal stability line
for $m=0$ and critical Reynolds numbers for $m>0$ for $\hat\eta=0.78$
and ${\rm Fr}=0.5$. The solid vertical line marks the
$\hat\mu=\hat\eta^2$ limit, the dashed vertical line marks  $\hat\mu=\hat\eta$  and the dotted vertical line marks $\hat\mu=\hat\eta^{1.5}$.}
\label{remu} 
\end{figure}
\begin{figure}[ht]
\psfig{figure=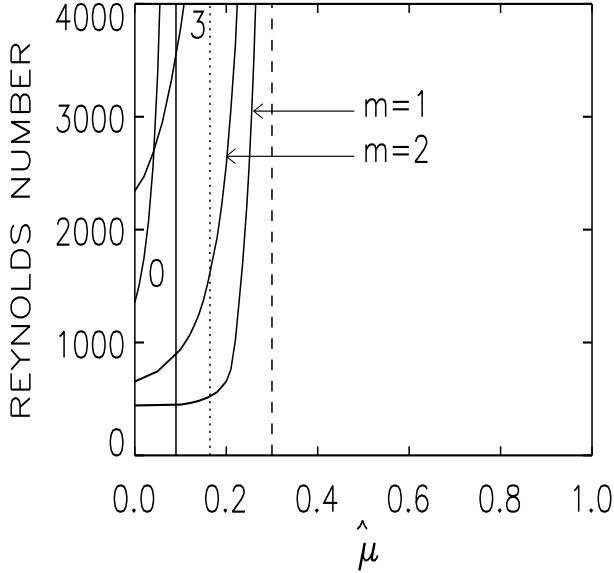,width=9.0cm,height=9.0cm} 
\caption{\label{remu2} The same as Fig.~\ref{remu}
but for a wide gap with $\hat\eta=0.3$.}
\end{figure}
\begin{figure}[ht]
\hbox{
\psfig{figure=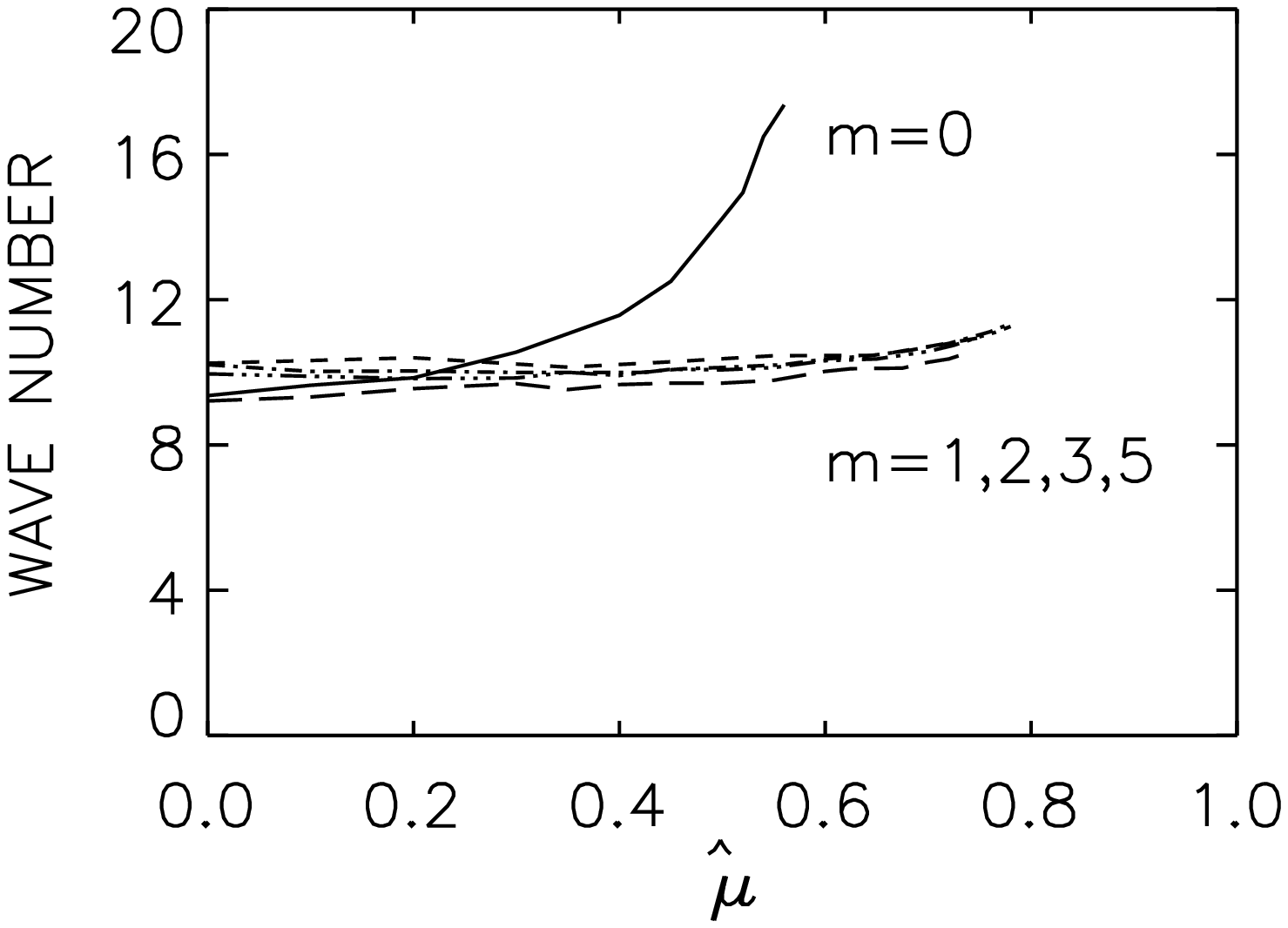,width=4.5cm,height=6.0cm}
\psfig{figure=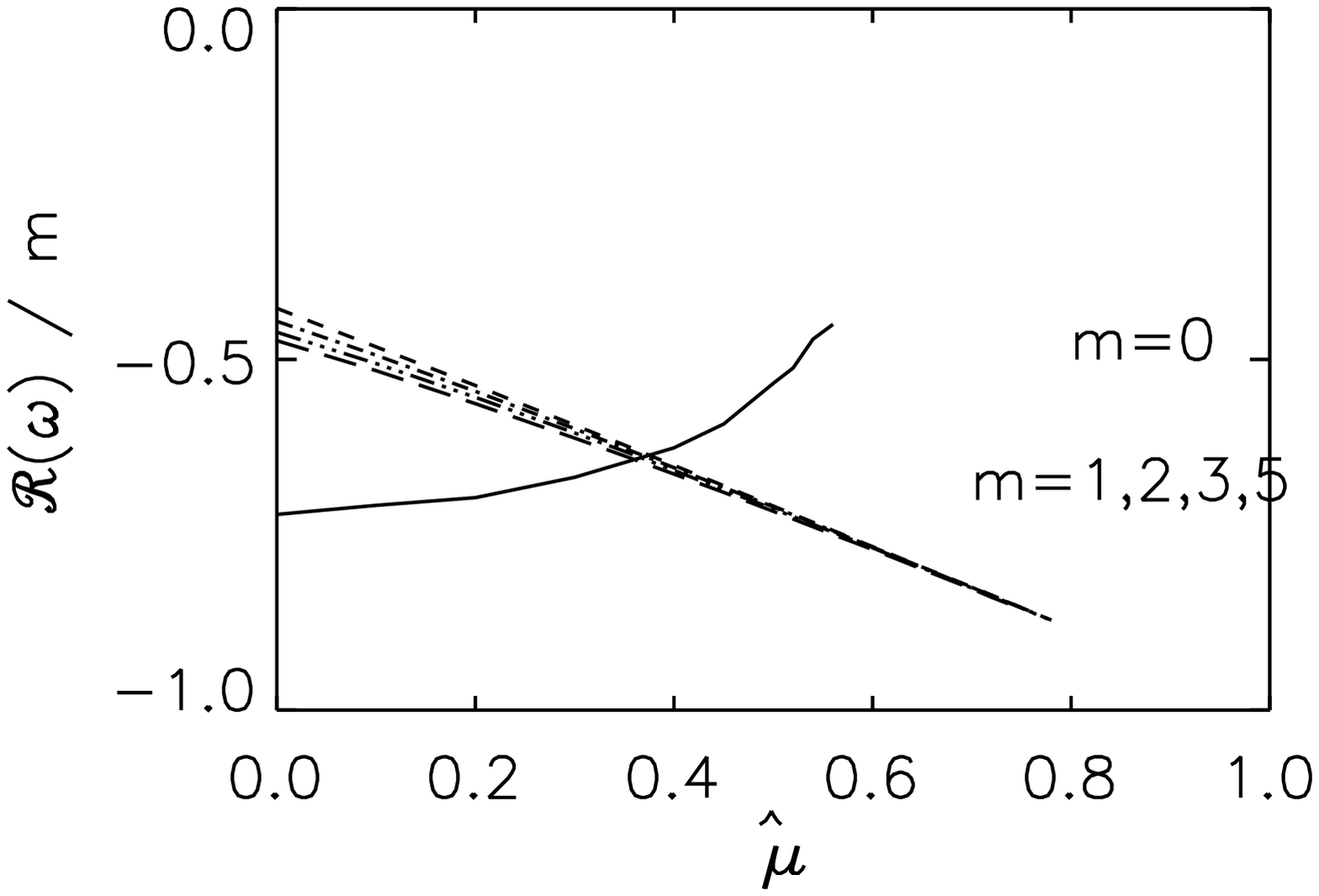,width=4.5cm,height=6.0cm}}
\caption{\label{kom} The same as Fig.~\ref{remu}
but for vertical wave number (left) and the pattern speed 
${\Re(\omega)}/m$ for   $m>0$ (right).}
\end{figure}

The dependence of the critical Reynolds numbers on $\hat\mu$ is given by
 Fig.~\ref{remu} for a narrow gap and in Fig. \ref{remu2} for a wide gap.
The exact line of marginal stability is plotted only for $m=0$. The
axisymmetric disturbances are unstable only for $\hat\mu < \hat\eta^2$
in accordance to the Rayleigh condition (\ref{ray}). For $m>0$ the slightly
unstable lines with $\Im(\omega)=-10^{-3}$ are given.
 
The nonaxisymmetric disturbances are unstable also beyond the Rayleigh line
(plotted as solid in the figures). 
The higher the $m$, however, the more the corresponding  instability line
approaches the Rayleigh line. 
Note, therefore, that the `stratorotational instability' (SRI, Dubrulle et al. 2004)
only produces low-$m$ modes. This is an indication that indeed within the
short-wave approximation (high-$m$) it does not exist (see R\"udiger, Arlt \&
Shalybkov 2002). 

For nonstratified Taylor-Couette flows the nonaxisymmetric modes are only the  most unstable disturbances    for counter-rotating cylinders
(e.g. Drazin \& Reid 1981). The
nonaxisymmetric instability of the stratified Taylor-Couette flow beyond the
Rayleigh line ($\hat\mu > \hat\eta^2$) leads to the existence of  some
critical value, $\hat\mu_c$, beyond which the nonaxisymmetric disturbances
are the most unstable. Our results show that $\hat \mu_c \sim 0.27$ and
almost independent of the Froude number for $\hat\eta=0.78$ (small gap) 
and $\hat\mu_c < 0$ for $\hat\eta=0.3$ (wide gap). It is possible that for some
$\hat\eta$  the nonaxisymmetric disturbances are the most unstable
for all values of $\hat\mu$ corresponding to unstable flows. 

There is another observation with the Figs. \ref{remu} and \ref{remu2}.
Approaching the line  $\hat\mu = \hat\eta$ the instability lines become more
and more steep. We did not find any solution for  $\hat\mu > \hat\eta$. 
If this is true one has to apply rather steep rotational profiles (not so steep
as for nonstratified fluids but also not so weak as for the MRI) to find  the modes of the linear SRI.
This result seems to be of relevant 
for the discussion of the stability or instability of Kepler disks.

\begin{figure}[ht]
\hbox{
\psfig{figure=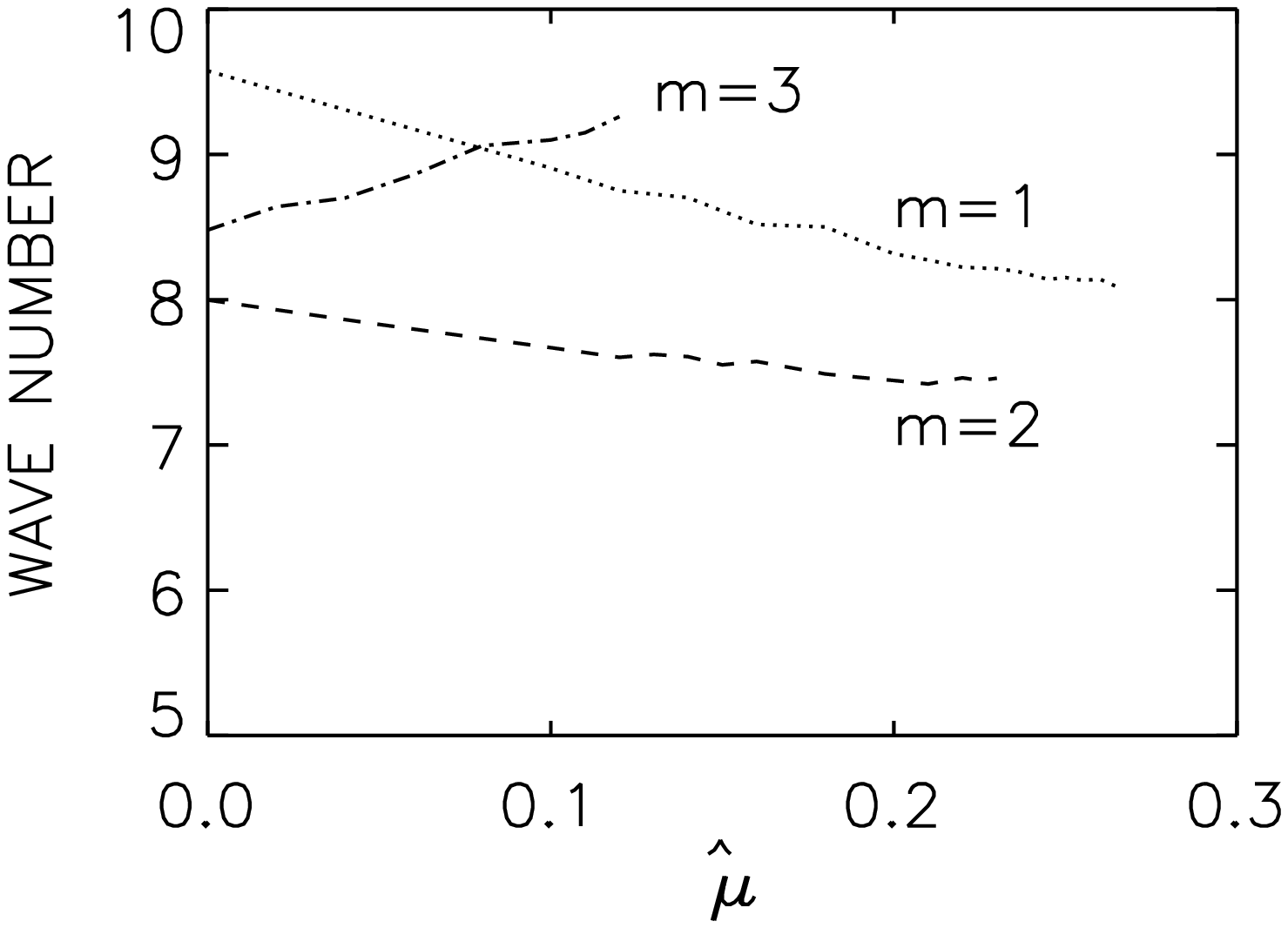,width=4.5cm,height=6.0cm}
\psfig{figure=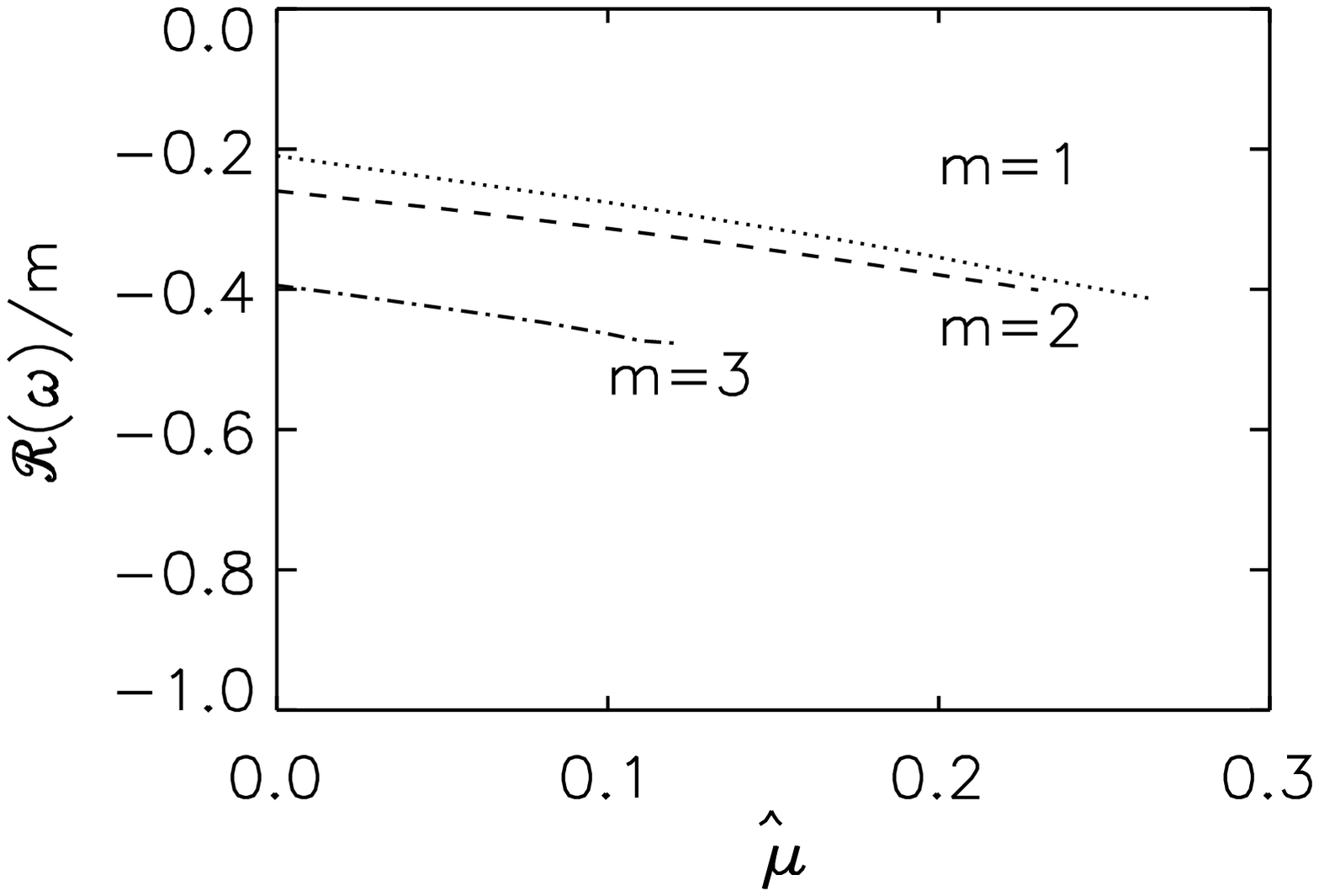,width=4.5cm,height=6.0cm}}
\caption{\label{kom03} The same as Fig.~\ref{kom} but for $\hat\eta=0.3$.}
\end{figure}

For the narrow gap ($\hat\eta=0.78$) the critical
Reynolds numbers of the nonaxisymmetric modes only slightly depend on $m$. 
The same is true for the critical vertical wave number and the pattern speed $\Re({\omega})/m$ (Fig.~\ref{kom}). The vertical wave number  only weakly 
 depends on $\hat\mu$ and the  values $\Re{(\omega)}/m$
 linearly run with $\hat\mu$.
The situation is  changed, however,  for the wide gap ($\hat\eta=0.3$).
All parameters now strongly depend on both $m$ and $\hat\mu$ (Fig.~\ref{kom03}). The trend for the vertical wave numbers  is opposite
for $m=3$ to those for $m=1$ and $m=2$.

The vertical wave numbers for both containers  are of order 10 for $m=1$. With our normalization the vertical extent of the Taylor vortices  is given by
\beg
\frac{\delta z}{R_{\rm out}}=
\frac{\pi}{k}\sqrt{{\hat\eta}({1-\hat\eta})}.
\ende
With the mentioned value of $k$ it is order of 0.1 for both the  small-gap case and the wide-gap case. The cell becomes thus rather  flat.  For nonstratified TC-flows one finds $\delta z \simeq R_{\rm out} - R_{\rm in}$ while the cells under the influence of an axial magnetic field become more and more prolate. 
 The stratification generally reduces
the height of the Taylor vortices.

Unlike to the nonstratified Taylor-Couette flow
the $\Re(\omega)$ is not zero for stratified Taylor-Couette 
even for axisymmetric disturbances ($m=0$). The onset of instability
is thus oscillatory (`overstability').  The question is whether  a 
 critical Froude number exists corresponding to the transition
from stationary solutions to oscillating solutions?
The answer is No.  One cannot   fulfill
 Eq.  (\ref{sysf}) for marginal stability ($\Im(\omega)=0$) without a finite 
real part of $\omega$ for $N^2 \ne 0$.  The axially stratified
Taylor-Couette flow bifurcates from the purely azimuthal flow through
a direct Hopf bifurcation to a wavy regime. Depending on the value
of $\hat \mu$ and $\hat \eta$
this new  regime can be either oscillating and axisymmetric or nonaxisymmetric and azimuthally drifting 
(see Figs. \ref{remu}, \ref{remu2}). For both our containers  the pattern speeds are negative for positive $m$. The drift of the spirals is thus always in the direction of the cylinder rotation.

Experiments have really demonstrated 
the oscillating  onset of the axisymmetric instability
(e.g. Caton, Janiaud \& Hopfinger 2000). 
It would be interesting to design   experiments with either
rotating outer cylinder or wider gap to probe the bifurcation from  the overstable 
oscillating axisymmetric flow pattern  to the spiral  nonaxisymmetric flow pattern.

\begin{figure}[ht]
\vbox{
\psfig{figure=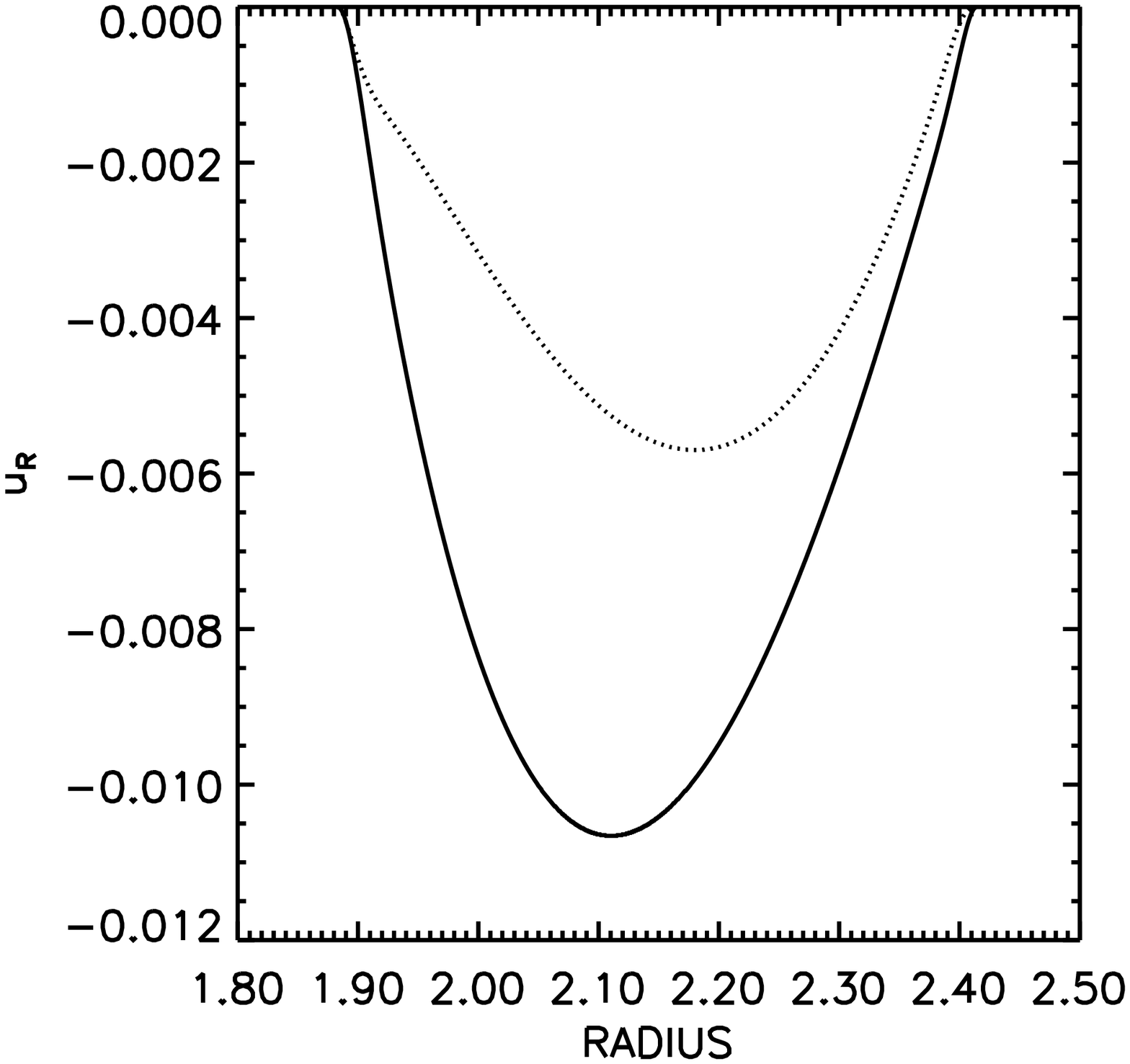,width=8cm,height=4.0cm}
\psfig{figure=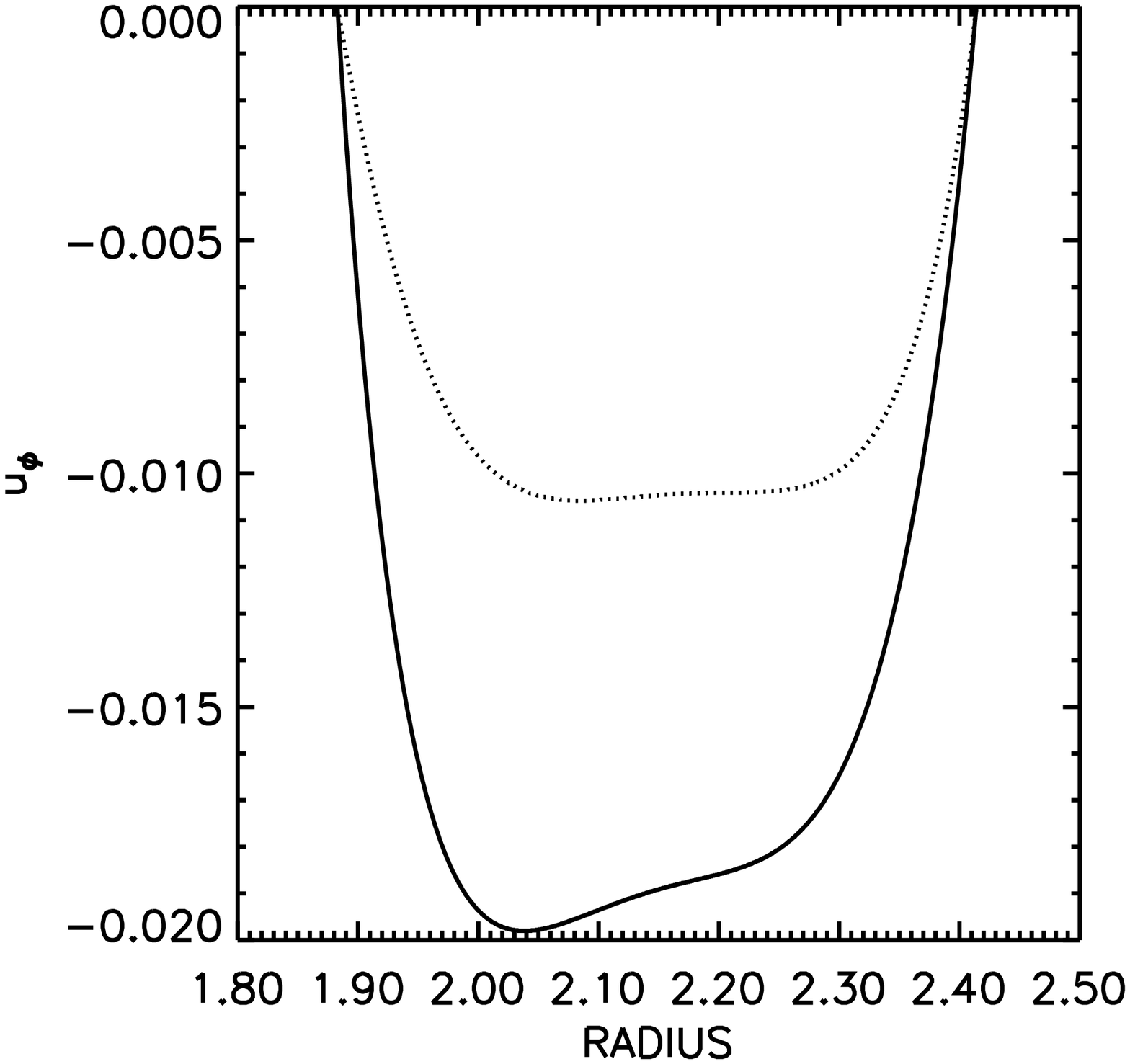,width=8cm,height=4.0cm}
\psfig{figure=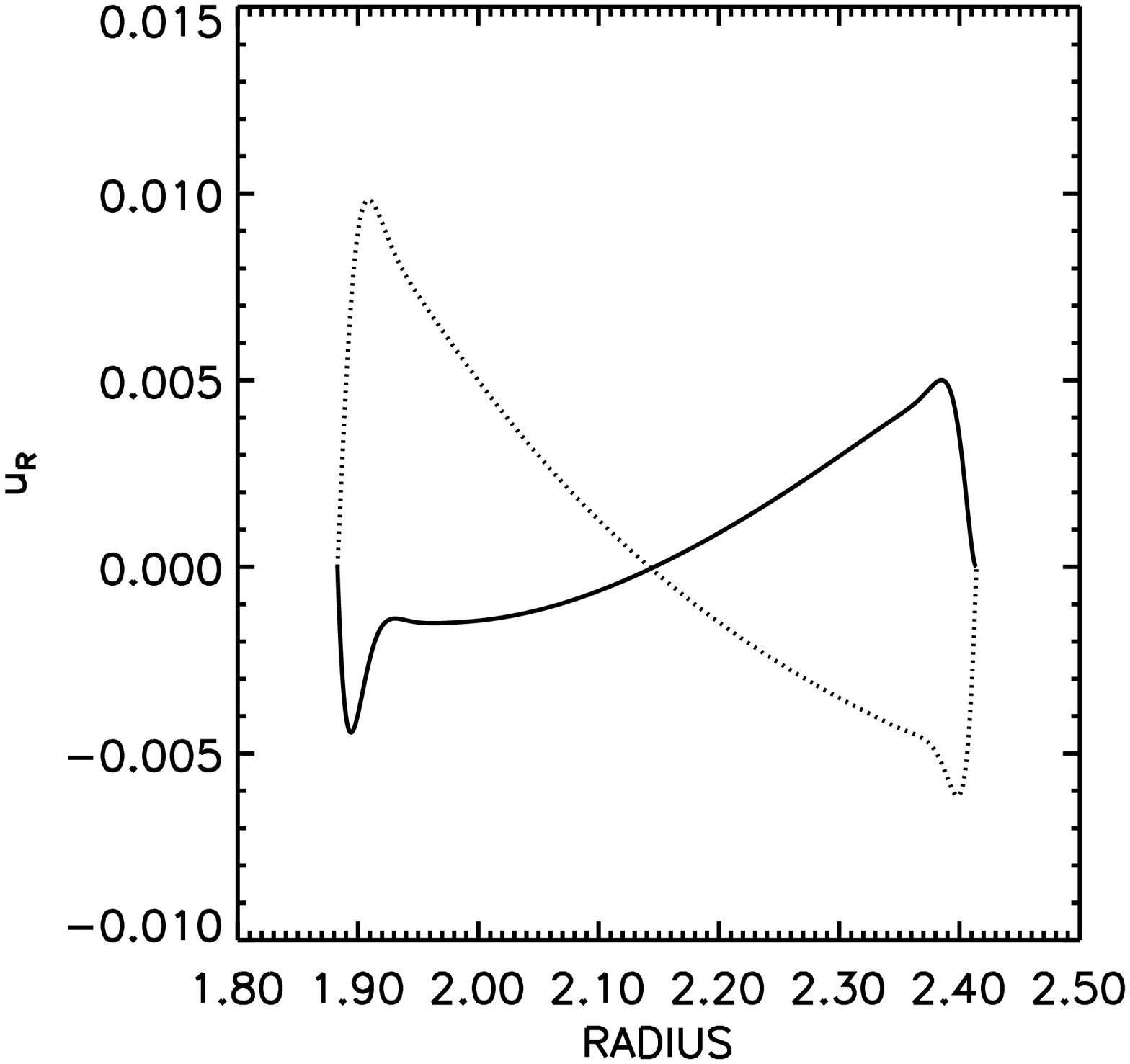,width=8cm,height=4.0cm}
}
\caption{\label{eign} The velocity eigenfunctions for $m=1$, $\hat\eta=0.78$,
$\hat\mu=0.7$, ${\rm Fr}=0.5$ with  Reynolds number, vertical wave number and 
$\Re(\omega)$. The dotted lines are the real part and solid lines are the
imaginary part.}
\end{figure}

As an example for the container with the narrow gap in Fig.~\ref{eign}
  the velocity eigenfunctions are  presented 
for $m=1$ and for $\hat\mu=0.7$ exceeding the value of  $\hat\eta^2$. The functions
are smooth enough and do not suggest that the instability
must be explained as a  boundary
effect. 
\begin{figure}[ht]
\vbox{
\psfig{figure=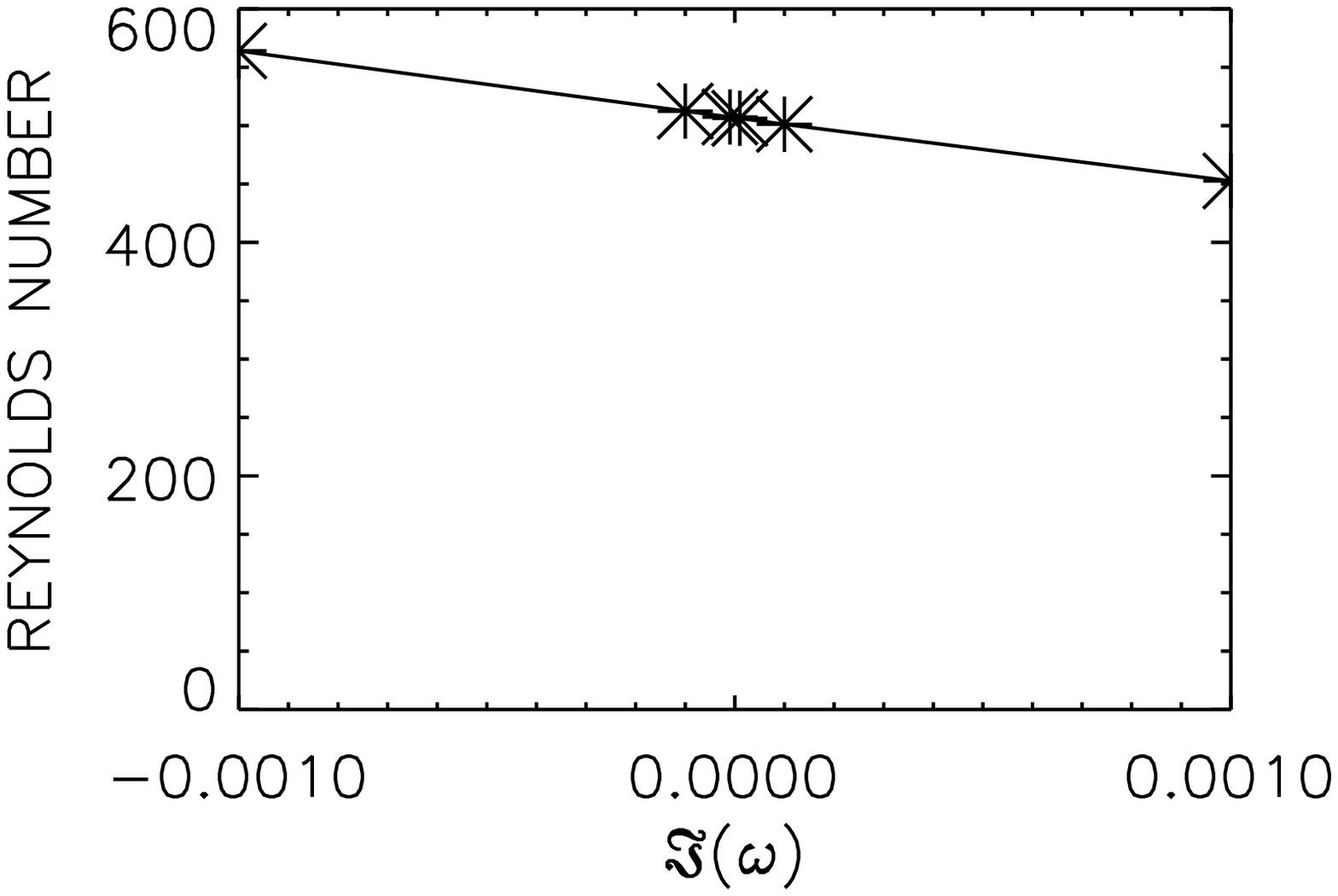,width=8cm,height=4.0cm}
\psfig{figure=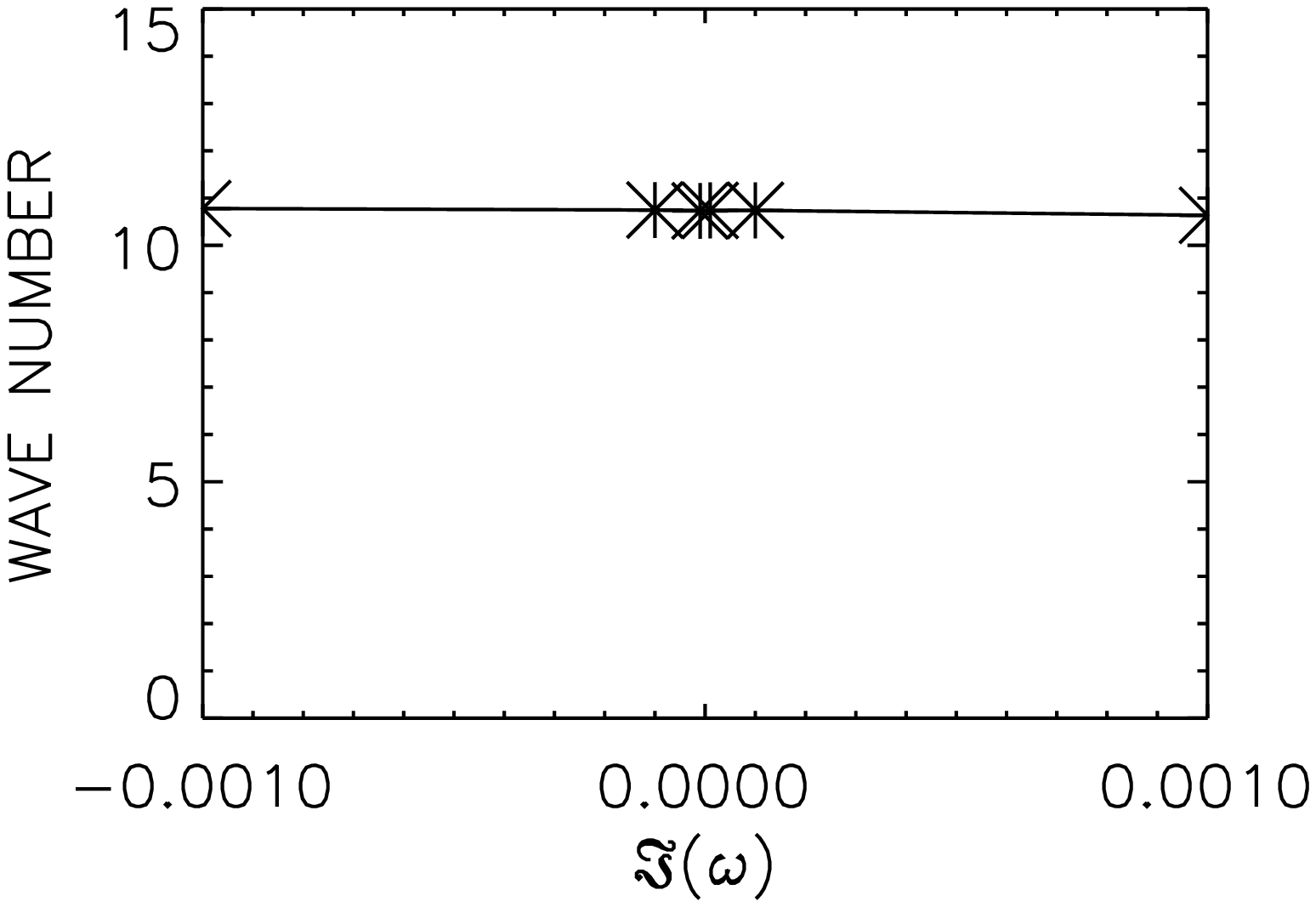,width=8cm,height=4.0cm}
\psfig{figure=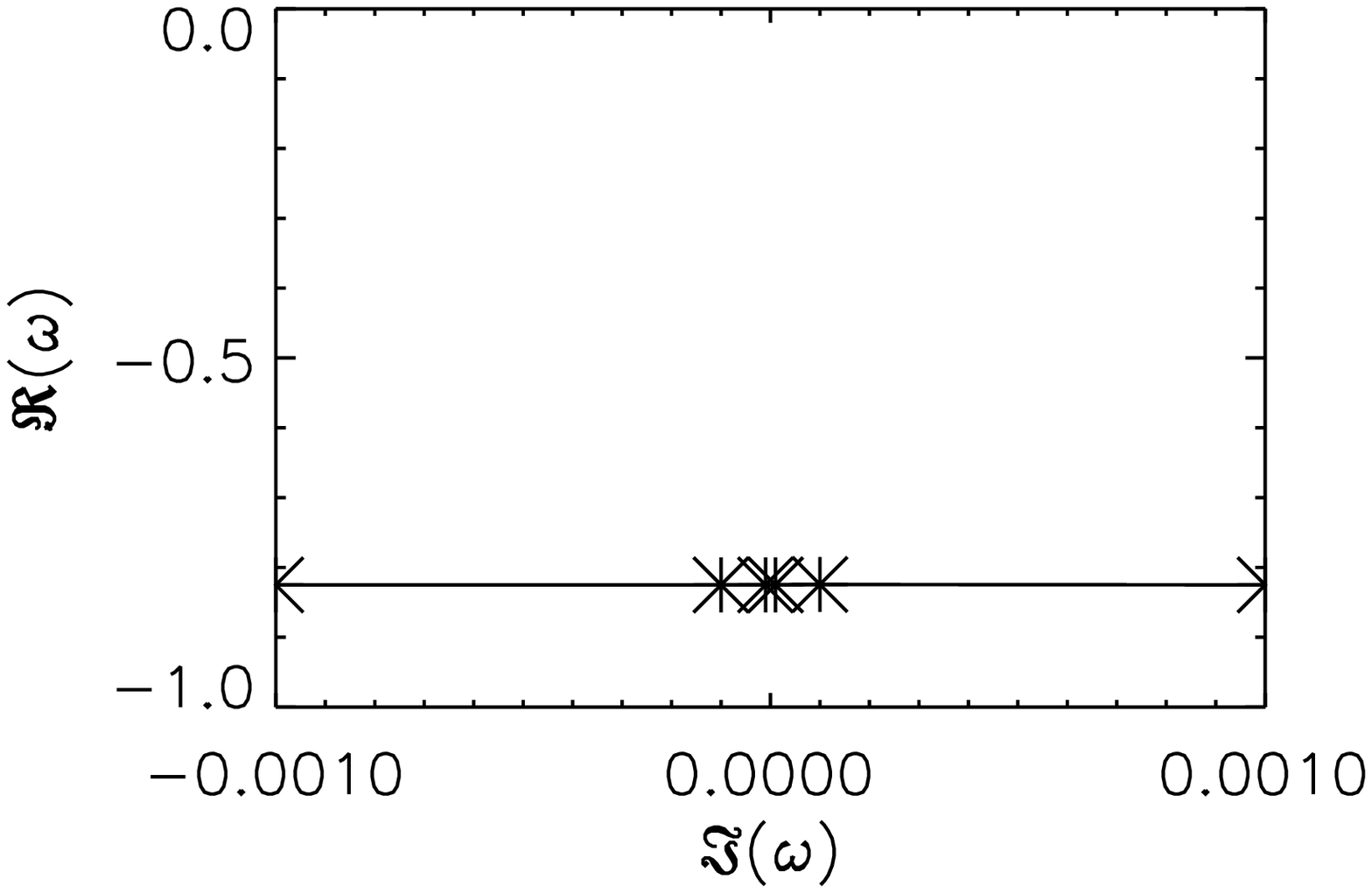,width=8cm,height=4.0cm}
}
\caption{\label{trans} The Reynolds number, vertical wave number and $\Re(\omega)$
 in the transition  from positive to negative
 $\Im(\omega)$ for $m=1$, $\hat\eta=0.78$, $\hat\mu=0.7$, 
${\rm Fr}=0.5$.}
\end{figure}

The Reynolds
number, the vertical wave number and the real part of $\omega$
for the transition  from positive to negative
imaginary part of $\omega$ are given in Fig.~\ref{trans}, i.e. for the transition from negative to positive growth rates. We find a  continuous transition across the marginal
stability line. It should thus be  possible to realize
the transition from stable to unstable flows in experiments. The
vertical wave number and real part of $\omega$ are hardly influenced by the transition 
 but, not surprisingly, the  Reynolds numbers has a remarkably  clear
trend. 
\section{Discussion}
\label{disc}
It is  shown that    the  Boussinesq approximation  yields  nonaxisymmetric disturbances with low $m$ 
of the stratified Taylor-Couette flow as   unstable even beyond the Rayleigh line $\hat\mu > \hat\eta^2$.  Our results, however, also show
that the critical Reynolds numbers are extremely increasing approaching  the
line $\hat\mu=\hat\eta$ so that as the condition for instability now the relation
\beg
\hat\mu<\hat\eta
\label{eta}
\ende
seems to appear  rather than $\hat\mu < 1$ according to 
Yavneh, McWilliams \& Molemaker (2001) and  Molemaker,
McWilliams \& Yavneh (2001). It is challenging to interprete the line $\hat\mu=\hat\eta$ with the (galactic) rotation profile $u_\phi=$const in the same sense as to interprete the line  $\hat\mu=\hat\eta^2$  with the rotation law for uniform specific angular momentum $R^2\Om=$ const. Below we shall consider the line 
 $\hat\mu=\hat\eta^{1.5}$ as concerning   the Kepler flow.\footnote{note, however, the general difference of  the Kepler rotation law $\Om\propto R^{-1.5}$ and the Taylor-Couette rotation law (\ref{Om})} Dubrulle et al. (2004) for their model of rotating plane Couette flow have  found unstable solutions also beyond 
the line $\hat\mu=\hat\eta$ but also in these computations the rotation profile must be steeper  than $R^{-2/3}$ (their Fig. 7). 

The  SRI also leads to the situation that
nonaxisymmetric disturbances can be the most unstable modes not only
for counterrotating cylinders but also  for corotating cylinders.
The characteristic values of  $\hat\mu$  where the
nonaxisymmetric disturbances are the most unstable ones strongly
depend on the gap width.  For
$\hat\eta=0.3$  all positive $\hat\mu$ are concerned (see Fig. \ref{remu2}). It cannot be  excluded that the nonaxisymmetric disturbances are the most unstable ones for all  $\hat\mu$
for $\hat\eta$ smaller than some critical value.

These results were obtained with the Boussinesq
approximation  so that   two restrictions remain. The vertical
density stratification should be weak enough and  the rotation should be so slow that the
centrifugal acceleration can be neglected in total. If one of these conditions is violated the 
Boussinesq approximation cannot be used  and the situation becomes much more
complicated. 
The  disagreement between the calculated
and the observed  critical Reynolds numbers for small
Fr (see Fig. \ref{comp}) may already indicate the violation of the Boussinesq
approximation  for strong stratifications. 

We have shown that for not too small negative  ${\rm d}\Om/{\rm d}R$ Taylor-Couette flows with vertical  density stratification become unstable against nonaxisymmetric disturbance with $m=1$ even if they are stable without density stratification. Kepler flows seem to be concerned by this phenomenon. In the Figs.  \ref{remu}, \ref{remu2}  the dotted lines represent the limit  $\hat\mu=\hat\eta^{1.5}$ which might   mimic the radial shear in Kepler disks. 
In both Figures  we find a critical Reynolds number of only about 500 for the lowest ($m=1$) mode. This is indeed a rather  small number whose meaning, however, should not be overestimated. Approaching the line 
 $\hat\mu=\hat\eta$ also these values become more and more large. Even more important is the finding that in  magnetohydrodynamic Taylor-Couette experiments the MRI only needs {\em magnetic} Reynolds numbers of O(10). This leeds to hydrodynamic Reynolds numbers exceeding  O($10{^6}$) only  for experiments  with liquid metals in terrestrial laboratories. For hot plasma with magnetic Prandtl numbers of order 10  (Noguchi \& Pariev 2003) the necessary hydrodynamic Reynolds number is also only O(1) or even smaller!

The existence  of the SRI might be important for astrophysical applications. As suggested  first by Richard \& Zahn (1999) one should not forget (in particular for protoplanetary disks) to probe hydrodynamical instabilities as the source for the necessary turbulence in accretion disks.  In partial confirmation of results of Dubrulle et al. (2004) we have here demonstrated that for the rotation laws of Taylor-Couette flows 
under the presence of vertical density gradients indeed linear hydrodynamic instabilities for low $m$ exist.
 The properties of these modes are described above. Whether they are important for the accretion disk physics is still an open question. Note that the density stratification in accretion disks completely vanishes in the equatorial region and that the unstable modes discussed above would more or less lead to a nonaxisymmetric structure of the disk rather than to turbulence. The next step in this direction must include the numerical simulation for  well-designed but simplified {\em global} accretion disk models.
  
\begin{acknowledgements}
D.S. thanks for financial support from the AIP visiting program.
His work was also partly supported by RFBR (grant 03-02-17522). GR thanks J.-P. Zahn for detailed discussions about the subject of this paper.
\end{acknowledgements}

\end{document}